\documentclass[%
twocolumn,
superscriptaddress,
showpacs,
 amsmath,amssymb,
 aps,prx,
]{revtex4-1}
 
\usepackage{graphicx}
\usepackage{dcolumn}
\usepackage{bm}
\usepackage{color}
\usepackage{multirow}
\usepackage{ulem}
\usepackage{hyperref}

\newcommand{\red}[1]{\textcolor{red}{#1}}

\definecolor{green}{rgb}{0,0.7,0.3}
\newcommand{\green}[1]{\textcolor{green}{#1}}

\newcommand{\andrew}[1]{\green{#1}}
\newcommand{\ket}[1]{|#1\rangle}

\begin{document}

\title{Restricted-Boltzmann-Machine Learning \\ for Solving Strongly Correlated Quantum Systems}

\author{Yusuke Nomura}
\email{nomura@ap.t.u-tokyo.ac.jp}
\affiliation{Department of Applied Physics, The University of Tokyo, 7-3-1 Hongo, Bunkyo-ku, Tokyo 113-8656, Japan}
\author{Andrew S. Darmawan}
\affiliation{Department of Applied Physics, The University of Tokyo, 7-3-1 Hongo, Bunkyo-ku, Tokyo 113-8656, Japan}
\author{Youhei Yamaji}
\affiliation{Department of Applied Physics, The University of Tokyo, 7-3-1 Hongo, Bunkyo-ku, Tokyo 113-8656, Japan}
\affiliation{JST, PRESTO, 7-3-1 Hongo, Bunkyo-ku, Tokyo 113-8656, Japan}
\author{Masatoshi Imada}
\affiliation{Department of Applied Physics, The University of Tokyo, 7-3-1 Hongo, Bunkyo-ku, Tokyo 113-8656, Japan}

\date{\today}

\begin{abstract}
    We develop a machine learning method to construct accurate ground-state wave functions of strongly interacting and entangled quantum spin as well as fermionic models on lattices. 
A restricted Boltzmann machine algorithm in the form of an artificial neural network is combined with a conventional variational Monte Carlo method with pair product (geminal) wave functions and quantum number projections. 
The combination allows an application of the machine learning scheme to interacting fermionic systems. 
The combined method substantially improves the accuracy beyond that ever achieved by each method separately, in the Heisenberg as well as Hubbard models on square lattices, thus proving its power as a highly accurate quantum many-body solver.
\end{abstract}

\maketitle
\section{Introduction}
Obtaining accurate ground-state wave functions of many-body quantum Hamiltonians is one of the grand challenges in condensed matter physics.
Great successes so far are, for example,  Bardeen-Cooper-Schrieffer (BCS) wave functions for conventional superconductivity~\cite{PhysRev.108.1175}, Bethe-ansatz wave function for one-dimensional interacting systems~\cite{PhysRevLett.20.1445}, 
and Laughlin wave functions for fractional quantum Hall effect~\cite{PhysRevLett.50.1395}.

However, in order to construct the ground-state wave functions of many-body interacting systems and grasp the essential physics
encoded in them,
we often need to resort to numerical estimates.
Currently, many numerical 
techniques are available such as 
the variational Monte Carlo (VMC) method~\cite{PhysRev.138.A442,PhysRevB.16.3081,doi:10.1143/JPSJ.56.3582,doi:10.1143/JPSJ.56.1490,PhysRevB.64.024512,doi:10.1143/JPSJ.77.114701},
the density matrix renormalization group~\cite{PhysRevLett.69.2863,PhysRevB.48.10345}, 
tensor network methods~\cite{doi:10.1080/14789940801912366,ORUS2014117}, 
and the path-integral renormalization group~\cite{doi:10.1143/JPSJ.69.2723}. 
Among them, the VMC method offers an accurate ground-state wave function for quantum spins as well as fermions on various lattices.

Recently, alternative approaches, based on machine learning, 
have attracted growing attention in many-body physics~\cite{Nphys_Carrasquilla,Nieuwenburg_Nphys,PhysRevLett.118.216401,doi:10.7566/JPSJ.85.123706,doi:10.7566/JPSJ.86.044708,doi:10.7566/JPSJ.86.093001,PhysRevE.95.062122,doi:10.7566/JPSJ.86.063001,PhysRevB.95.245134,Mehta_arXiv,PhysRevE.96.022140,PhysRevB.96.184410,PhysRevB.95.035105,PhysRevLett.119.030501,Sci_rep_Broecker,Broecker_second_arXiv,PhysRevX.7.031038,Ch'ng_second_arXiv,Morningstar_arXiv,Biamonte-nature,Millis_arXiv,Bukov_arXiv,PhysRevB.94.165134,PhysRevB.95.241104,PhysRevB.96.041119,PhysRevB.96.161102,PhysRevE.96.051301,Torlai_arXiv2,Costa_arXiv,Fujita_arXiv,PZhang_arXiv,WJ_Rao_arXiv,Yoshioka_arXiv,SaitoKato_arXiv}.
In particular, Carleo and Troyer~\cite{Carleo602} have proposed a machine-learning algorithm, which uses a restricted Boltzmann machine (RBM) as a variational wave function
$| \Psi \rangle$ for representing the ground states of quantum spin systems. 
In this scheme, hidden artificial neurons are introduced on top of the physical degrees of freedom (quantum spins), to mediate entanglement in the state.
The RBM variational wave functions
are self-optimized through 
machine learning.


We can express a general quantum state
$| \Psi \rangle$ 
by using the Fock space basis $\{|x\rangle\}$ in the form of a variational function as
\begin{eqnarray}
| \Psi \rangle = \sum_x |x\rangle \mathcal{F}(x) \langle x | \phi_{\rm ref} \rangle
 \label{eq.PsiFphi}
\end{eqnarray}
with a correlation factor $\mathcal{F}(x)$~\cite{PhysRev.98.1479,PhysRevLett.10.159} and a reference state $| \phi_{\rm ref} \rangle$. 
The RBM wave function in Ref.~\cite{Carleo602} is obtained by employing
an RBM for $\mathcal{F}(x)$ 
and a product state for $| \phi_{\rm ref}\rangle$ so that $\langle x| \phi_{\rm ref}\rangle=1$ is satisfied for any orthonormalized complete set $x$. 
The product state
$| \phi_{\rm ref} \rangle$ is not able to describe nonlocal quantum entanglement, although it is essential in strongly correlated systems.
Then the entanglement has to be represented solely by the RBM factor $\mathcal{F}(x)$.
However, alternative choices of $| \phi_{\rm ref} \rangle$ 
may already incorporate typical quantum correlations and  
can potentially allow $|\Psi \rangle$ to more efficiently capture ground state entanglement. 
In fact, in the many-variable VMC (mVMC) method~\cite{doi:10.1143/JPSJ.77.114701},
a pair-product (PP) wave function (or 
 equivalently geminal wave function in quantum chemistry~\cite{Hurley1953,Rassolov2002,Shao2006})
is chosen as $| \phi_{\rm ref} \rangle$, which can efficiently capture a substantial part of the non-local entanglement in strongly correlated quantum systems by using 
many variational parameters.

In this paper, we propose a variational wave function for studying strongly correlated quantum systems called RBM+PP which combines flexible and nonempirical correlation factor given by RBM and 
entangled reference state given by the PP wave function to 
inherit the advantages of both. When applied to the  two-dimensional (2D) Heisenberg model on a square lattice, we show that our method significantly outperforms the original RBM method ~\cite{Carleo602}, which itself outperforms existing numerical techniques for finite lattices based on tensor networks.

The PP wave function can also flexibly incorporate non-local correlations in fermionic systems and account for the fermionic sign, allowing RBM+PP to be applied to interacting systems of fermions. When applied to the Hubbard model we show that the combined method achieves greater accuracy than either method applied separately. 
To the best of our knowledge, this is the first application of the RBM-based wave functions to interacting fermions.
The RBM+PP method thus provides a powerful tool not only for quantum spins but also for highly entangled quantum states such as strongly correlated itinerant fermions.

The structure of the paper is as follows. 
In Sec.~\ref{sec_method}, we explain the RBM+PP method after introducing individual RBM and PP wave functions. 
In Sec.~\ref{sec_results}, we apply the RBM+PP scheme to Heisenberg and Hubbard models and show significant improvement from RBM and mVMC results.   
The representability of the RBM+PP wave function is discussed in detail in Sec.~\ref{sec_discussion}. 
Finally, we give a summary and present future perspectives in Sec.~\ref{sec_summary}. 
 
\section{RBM+PP method}
\label{sec_method}
In this section, we will define RBM and PP states and explain how these are combined in RBM+PP method.

\subsection{RBM wave function}
\label{sec_RBM_wf}
\begin{figure}[tb]
\vspace{0cm}
\begin{center}
\includegraphics[width=0.48\textwidth]{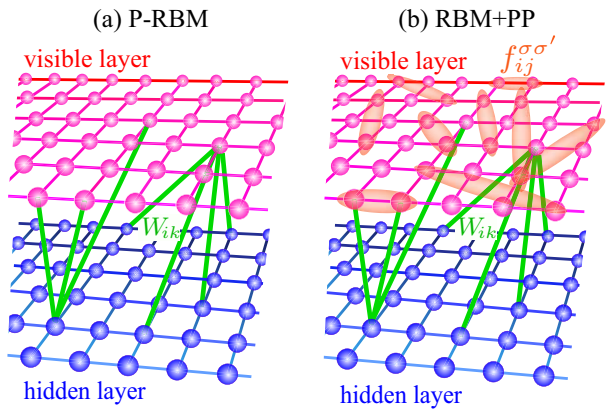}
\caption{
Schematic illustration of (a) P-RBM and (b) RBM+PP to represent many-body wave function. 
Physical variables in the visible layer couple to artificial neurons in the hidden layer through $W_{ik}$ interactions in Eq.~(\ref{eq:Psi(x)}).
Whereas no entanglement among physical variables exists in the absence of the hidden layer in P-RBM, RBM+PP provides direct entanglement via 
$f_{ij}^{\sigma \sigma'}$ parameters in Eq.~(\ref{eq_PP_state}).
For visibility, only a small portion of connections by $W_{ik}$ and $f_{ij}^{\sigma \sigma'}$ are shown. 
}
\label{fig_RBM}
\end{center}
\end{figure}

The RBM state in Ref.~\cite{Carleo602} for spin Hamiltonians is given by setting ${\mathcal F}(x) = {\mathcal N}(x)$ with a neural-network correlation factor ${\mathcal N}(x)$ and  
 $| \phi_{\rm ref}\rangle$ to be the product state $| \phi_{\rm product} \rangle$ ($\langle x | \phi_{\rm product} \rangle = 1$) in Eq.(\ref{eq.PsiFphi}), which from now we refer to as product-basis RBM (P-RBM) (Fig.~\ref{fig_RBM}(a)). 
 $\mathcal{N}(x)$ is defined by an artificial neural network (ANN) as 
\begin{eqnarray}
  \mathcal{N}(x)  \! = \! \sum_{ \{ h_k\}}  \exp \Bigl (   \sum_{i} a_i \sigma_i  + \sum_{i,k} W_{ik}  \sigma_i h_k +\sum_{k} b_k h_k \Bigr ),  
\label{eq:Psi(x)}
\end{eqnarray}
where $x = (\sigma_1, \sigma_2, \ldots, \sigma_{N_{\rm visible}})$ is a real space configuration of $N_{\rm visible}$ physical variables
and $\sigma_i$ is the $i$ th discrete-valued physical variable (visible-layer spin variable).
In the $S=\frac{1}{2}$ Heisenberg model, we take $\sigma_i = 2 S_i^z = \pm 1$ with $S_i^z$ being the $z$-component of the $S=\frac{1}{2}$ spin at site $i$. Here $N_{\rm visible}$ is equal to the number of sites $N_{\rm site}$.  The auxiliary pseudo spin variables
$h_k = \pm 1 $  are for the hidden neurons and $\{ a_i, W_{ik}, b_{k} \}$ is a set of variational parameters. 
In this study, we take variational parameters to be real. 
Importantly, as there are no weights connecting hidden neurons, the sum over hidden variables can be evaluated exactly and Eq.(\ref{eq:Psi(x)}) can be reduced to the form $\mathcal{N}(x) \equiv \prod_k 2 \cosh \Bigl ( b_k + \sum_i W_{ik} \sigma_i  \Bigr ) 
\times e^{\sum_i a_i \sigma_i }$, which can be computed efficiently for each $x$.
 
In fermionic models we define a different RBM state which we refer to as F-RBM. The correlation factor ${\mathcal F}(x) = {\mathcal N}(x)$ is 
taken similarly to, but by slightly modifying the spin case.  
For $\ket{\phi_{\rm ref}}$, our most primitive choice is a Fermi-sea state rather than the product state. 
This is because the product state is too poor in representing fermionic entanglement. 
The Fermi sea is 
 the ground state of the noninteracting fermion lattice models, much like product states are ground states of noninteracting spin models, and is thus able to account for the most primitive part of the fermionic entanglement and signs. 
We can use the same form of ${\mathcal N}(x)$ as Eq.~(\ref{eq:Psi(x)})
by doubling the number of the visible-layer variables ($N_{\rm visible} = 2 N_{\rm site}$) and
mapping fermionic modes to these spins as
$(\sigma_{2i},\sigma_{2i-1})=(2n_{i\uparrow} \!-\!1,2n_{i\downarrow} \!-\!1)$
where $n_{i\sigma}$ is the number operator for the fermions at site $i$ with spin $\sigma$.
 

 As discussed in Ref.~\cite{Carleo602}, the accuracy of the wave function can be controlled by the ``hidden variable density" $\alpha$, which is defined as $N_{\rm hidden}$/$N_{\rm visible}$ with the number of neurons $N_{\rm hidden}$ in the hidden layer and the number of  
physical variables $N_{\rm visible}$ in the visible layer.

\subsection{PP wave function}
Whereas the RBM states in the first step use the product and Fermi-sea states for $\ket{\phi_{\rm ref}}$, 
more sophisticated choices of $\ket{\phi_{\rm ref}}$ are able to incorporate more involved entanglement 
directly into the state. 
In this work we will use the pair-product (PP) state $| \phi_{\rm pair} \rangle$  as $\ket{\phi_{\rm ref}}$.  The PP wave function is given by 
\begin{equation}
| \phi_{\rm pair} \rangle = \Bigl (  \sum_{i, j=1}^{N_{\rm site}}  \sum_{\sigma, \sigma'=\uparrow,\downarrow } f_{i  j}^ {\sigma \sigma'} c^{\dagger}_{i \sigma} c^{\dagger}_{j \sigma'}   \Bigr )^{N_{\rm e}/2}|0\rangle
\label{eq_PP_state}
\end{equation}
where $N_{\rm e}$ is the number of electrons, $f_{i  j}^ {\sigma \sigma'}$ are variational parameters,
and $c_{i\sigma}^{\dagger}$ is the operator creating a $\sigma$-spin electron at site $i$.
For a given real space configuration $x$, $\phi_{\rm pair}(x)=\langle x | \phi_{\rm pair}\rangle$ can be expressed as the Pfaffian of a matrix, and can be calculated efficiently, much like $\mathcal{N}(x)$. 
Note that an accurate description of the node position is crucially important for fermionic wave function while the simple product state with the positive definite coefficients does not describe the node.  In contrast, the Pfaffian wave function is able to optimize the nodal structure within the framework of the Pfaffian wave function.
Therefore, the PP wave function can account for typical non-local entanglement not only in non-frustrated spin systems but also in frustrated spin and  fermionic systems. 

In spin models, 
to prohibit the double occupation, PP wave function is supplemented by 
the Gutzwiller factor ${\mathcal P}_{\rm G}^{\infty} = \prod_{i} (1- n_{i \uparrow} n_{i\downarrow})$.
This form of reference function  $| \phi_{\rm ref}\rangle = {\mathcal P}_{\rm G}^{\infty} | \phi_{\rm pair}\rangle$ is able to represent resonating valence bond (RVB) wave functions~\cite{ANDERSON1973153,ANDERSON1196}.

\subsection{RBM+PP wave function}
In this paper, we will study the combined wave function RBM+PP (Fig.~\ref{fig_RBM}(b)) with 
${\mathcal F}(x) = {\mathcal N}(x)$  and $\ket{\phi_{\rm ref}}=\ket{\phi_{\rm pair}}$ (itinerant fermions) or ${\mathcal P}_{\rm G}^{\infty} | \phi_{\rm pair}\rangle$ (spins) in Eq.~(\ref{eq.PsiFphi}).
We remark that the RBM+PP wave function is similar to that used in the
mVMC method~\cite{doi:10.1143/JPSJ.77.114701},
except that the mVMC method uses an empirical form of $\mathcal{F}(x)$ instead of more flexible and unbiased neural-network factor $\mathcal{N}(x)$. Specifically, in the mVMC method,  
$\mathcal{F}(x)$ is given by $\mathcal{F}(x) = \langle x | {\mathcal P}_{\rm G} {\mathcal P}_{\rm J} | x  \rangle $ with 
Gutzwiller ${\mathcal P}_{\rm G}$ (controlling 
double occupancy) \cite{PhysRevLett.10.159} and Jastrow ${\mathcal P}_{\rm J}$ (for  long-ranged charge-charge correlation) \cite{PhysRev.98.1479} factors.
In Appendix \ref{Sec.RBM_Gutzwiller}, we show that the neural-network factor $\mathcal{N}(x)$ is indeed more flexible than the empirical factors by showing that  $\mathcal{N}(x)$ can represent both the Gutzwiller and Jastrow factors. The $\mathcal{N}(x)$ factor can also represent many-body (more than two-body) correlations~\cite{PhysRevLett.45.573} at the same time.

Various symmetries can be imposed on the wave function to improve accuracy and reduce computational cost~\cite{Mizusaki}. 
In this study, as in Ref.~\cite{Carleo602}, we impose translational symmetry in the variational parameters $\{ a_i, W_{ik}, b_k \}$ in the RBM.
In the antiferromagnetic Heisenberg and half-filled Hubbard models, 
because $\sum_{i} \sigma_i = 0$ holds in the ground state, 
the translationally-invariant bias term $a_i = a$ becomes irrelevant. Therefore, we neglect it.
Furthermore, as the ground states of the Hamiltonians considered have total spin $S=0$ and momentum $K=0$, we apply the projections onto these subspaces, respectively, $\mathcal{L}^{S=0}$ and $\mathcal{L}^{K=0}$, to $\ket{\phi_{\rm pair}}$ or ${\mathcal P}_{\rm G}^{\infty} | \phi_{\rm pair}\rangle$
to improve accuracy. 
If we apply both $\mathcal{L}^{S=0}$ and $\mathcal{L}^{K=0}$, the reference state becomes $\ket{\phi_{\rm ref}}=\mathcal{L}^{K=0}\mathcal{L}^{S=0}\ket{\phi_{\rm pair}}$ for itinerant fermions and $\ket{\phi_{\rm ref}}=\mathcal{L}^{K=0}\mathcal{L}^{S=0} {\mathcal P}_{\rm G}^{\infty}  \ket{\phi_{\rm pair}}$ for spins. 

\subsection{Machine learning of variational parameters}
The form of the wavefunction in Eq. (\ref{eq.PsiFphi}) allows calculating physical quantities, and derivatives with respect to variational parameters to be approximated efficiently using Markov chain Monte Carlo sampling over the probability distribution $p(x)=\langle \Psi | x \rangle \langle x | \Psi \rangle/\langle \Psi | \Psi \rangle$. We use a machine learning method (called stochastic reconfiguration in Ref. \cite{PhysRevB.64.024512} and natural gradient in Refs. \cite{IEEE_Amari,neural_Amari}) to optimize the variational parameters in the wave function with respect to the energy. The computational cost of the optimization scales as ${\mathcal O} (N_{\rm site}^3)$ for RBM+PP, compared to  ${\mathcal O} (\alpha N_{\rm site}^2)$ for P-RBM~\cite{Carleo602}.
Thus, the improved accuracy of RBM+PP over P-RBM comes at some additional computational cost.
Details and comparisons of the variational wave functions we introduced in this section are listed in Appendix \ref{ListVF}.

\section{Models}
We apply  the RBM+PP scheme
to calculate the ground states of 2D $S=\frac{1}{2}$ antiferromagnetic (AFM) Heisenberg and 2D Hubbard models on the square lattice. Their Hamiltonians are defined as follows:
\begin{eqnarray}
 {\mathcal H}_{{\rm Heisenberg}} &=&  J \sum_{(i,j)} {\bm S}_i  \cdot {\bm S}_j
  \quad (J>0), \\
 {\mathcal H}_{{\rm Hubbard}} &=& - t \sum_{(i,j)\sigma}  c^{\dagger}_{i\sigma} c_{j\sigma} + U\sum_{i}  n_{i \uparrow} n_{i\downarrow}.  
\end{eqnarray}
The sum over sites $i$, $j$ is restricted to nearest-neighbor pairs. 
We take the exchange constant $J$ and hopping  $t$ as an energy unit in each case. 
The onsite repulsion $U$ controls the strength of correlation in the Hubbard model.

In our calculations, fully periodic boundary conditions are imposed for the Heisenberg model, while periodic (in $x$-direction) and anti-periodic (in $y$-direction) (P-AP) boundary conditions are imposed for the Hubbard model. 
Further details of the computation conditions are available in Appendix \ref{Details}.

\section{Results} 
\label{sec_results}
\subsection{Heisenberg model}
\begin{figure}[tb]
\vspace{0.0cm}
\begin{center}
\includegraphics[width=0.49\textwidth]{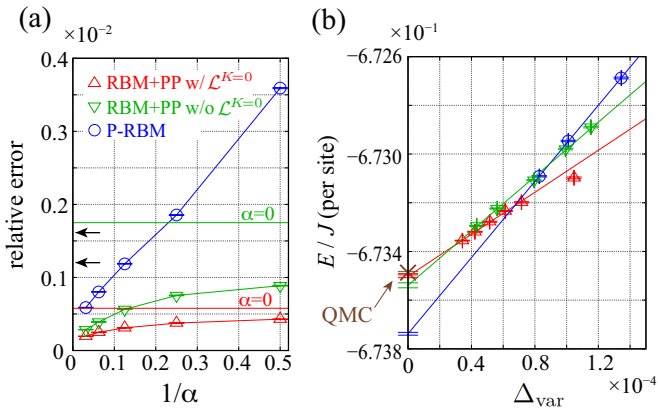}
\caption{ 
RBM+PP results for energy of 2D AFM Heisenberg model defined on $8\times 8$ square lattice with fully periodic boundary condition.  
(a) Relative error of energy to SSE-QMC energy ($E/J = -0.673487(4)$) \cite{PhysRevB.56.11678} as a function of $1/\alpha$ ($\alpha$: hidden variable density).
$\bigtriangleup$ ($\bigtriangledown$) symbol: RBM+PP $| \Psi \rangle = \mathcal{NL}^{K=0}  \mathcal{L}^{S=0} {\mathcal P}_{\rm G}^{\infty} | \phi_{\rm pair} \rangle $ ($| \Psi \rangle = \mathcal{NL}^{S=0} {\mathcal P}_{\rm G}^{\infty} | \phi_{\rm pair} \rangle$). 
$\bigcirc$ symbol:  P-RBM $| \Psi \rangle = \mathcal{N} | \phi_{\rm product} \rangle$.
$\alpha=0$ (solid horizontal lines) corresponds to the mVMC results (red: $| \Psi \rangle = \mathcal{L}^{K=0}  \mathcal{L}^{S=0} {\mathcal P}_{\rm G}^{\infty} | \phi_{\rm pair} \rangle$, green: 
$| \Psi \rangle =  \mathcal{L}^{S=0} {\mathcal P}_{\rm G}^{\infty} | \phi_{\rm pair} \rangle $).
Two arrows (from top to bottom) indicate the result of entangled-plaquette states (EPS)~\cite{1367-2630-11-8-083026} and 
	variational QMC to evaluate the projected entangled pair states (PEPS) for virtual bond dimensions of 16~\cite{PhysRevB.83.134421}. 
(b) Variance $\Delta_{\rm var}$ extrapolation of energy.
The cross ($\times$) on the ordinate shows the SSE-QMC energy. The data points plotted in this variance range are $\alpha =0$, 2, 4, 8, 16, 32 (from right to left) for 
red $\bigtriangleup$ symbols, 
$\alpha =2$, 4, 8, 16, 32 for 
green $\bigtriangledown$ symbols, 
and $\alpha =8$, 16, 32 for blue $\bigcirc$ symbols, respectively.
Linear fit and extrapolation to $\Delta_{\rm var}\rightarrow 0$
is shown as sold lines. 
All the data points in this range except $\alpha =0$ data in red 
are used in the fit.  
Error bars show standard errors of Monte Carlo measurements of energy (and also variance in case of (b)) for the optimized variational wave function.  
}
\label{fig_Spin1}
\end{center}
\end{figure}

Figure~\ref{fig_Spin1} shows  the RBM+PP ground-state energy for $8 \times 8$ 2D Heisenberg model compared to quantum Monte Carlo calculations using the stochastic series expansion (SSE-QMC) 
at sufficiently low temperature $T$ of  $1/T=64$~\cite{PhysRevB.56.11678}, which  gives practically exact ground state energy. 
For comparison, the mVMC results~\cite{doi:10.1143/JPSJ.77.114701} and the P-RBM wave function employed in Ref.~\cite{Carleo602} are shown. 
Here, mVMC is equivalent to $\alpha=0$ of RBM+PP. This is because the Gutzwiller factor $\mathcal{P}_{\rm G}$ is fixed to freeze charge degrees of freedom and Jastrow factor $\mathcal{P}_{\rm J}$ becomes irrelevant in the absence of charge degrees of freedom.

As discussed in Refs.~\cite{PhysRevLett.61.365,PhysRevB.42.6555}, the resonating valence bond (RVB) wave function is known to provide a highly accurate description of the 2D Heisenberg model. 
The relative error in the result of the mVMC function, which can represent the RVB wave function, is indeed less than 0.2 percent. 
We see that the non-empirical P-RBM wave function is also powerful, giving a comparable accuracy to the mVMC results. 
The RBM+PP wave functions, which take advantages of the above two, substantially improve the accuracy of independent mVMC and P-RBM schemes.

%

 It is interesting to note that  
  each of the three curves in Fig. ~\ref{fig_Spin1} corresponds to the results with the very same form of correlation factor $\mathcal{N}(x)$ but with different reference wave functions $| \phi_{\rm ref} \rangle$. 
The lowest energy is obtained when 
$\ket{\phi_{\rm ref}}=\mathcal{L}^{K=0}\mathcal{L}^{S=0}  {\mathcal P}_{\rm G}^{\infty}| \phi_{\rm pair} \rangle$ 
followed by $\ket{\phi_{\rm ref}}=\mathcal{L}^{S=0}  {\mathcal P}_{\rm G}^{\infty} | \phi_{\rm pair} \rangle$, with the product state $\ket{\phi_{\rm ref}} = \ket{\phi_{\rm product}} $ having the highest energy. Therefore, improving reference function helps the RBM to learn the ground state more efficiently. 

\begin{table} [tb]
\caption{ 
P-RBM and RBM+PP results for spin structure factor $S(\pi,\pi)$ in 2D Heisenberg model. 
For comparison, $S(\pi,\pi)/N_{\rm site} = 0.05986(3)$ in mVMC and $0.059280(3)$ in SSE-QMC \cite{PhysRevB.56.11678} results.
 }
\vspace{-0.5cm}
\begin{center}
\begin{tabular}{@{ \ }c@{ \ \ \ \ } c@{\ \ \  \  }  c   @{\ \ \ \  } c @{ \ }}
\hline 
\hline
\multirow{2}{*}{  \vspace{-0.1cm} wave function} &    \multicolumn{3}{c}{   $S(\pi,\pi) /N_{\rm site} \times 10^2  $ }  \\
\cline{2-4}
   & $\alpha= 2$ & $\alpha= 8$ & $\alpha= 32$ \\
\hline
$ \mathcal{N}  | \phi_{\rm product} \rangle $  
&\ 6.017(2)  & 5.955(2) & 5.946(2) \\
 $ \mathcal{NL}^{K=0}  \mathcal{L}^{S=0}  {\mathcal P}_{\rm G}^{\infty} | \phi_{\rm pair} \rangle $ 
 &\ 5.969(2)  &5.956(2) & 5.944(2) \\
\hline
\hline
\end{tabular}
\end{center}
\label{Tab_spin}
\end{table}

In Fig.~\ref{fig_Spin1}(b), we plot the total energy as a function of its variance $\Delta_{\rm var} \! \! = \! \! \bigl( \langle {\mathcal H}^2 \rangle \! - \! \langle {\mathcal H} \rangle ^2 \bigr) / \langle {\mathcal H} \rangle^2$.
The variance is zero in the case of an exact ground state (or more generally, an exact eigenstate of Hamiltonian). 
By the linear fit of the energy as a function of the variance and extrapolating to $\Delta_{\rm var} =0$,  we can obtain a more accurate estimate of the ground state energy~\cite{JPSJ2001.2287.2001,PhysRevB.58.6800,PhysRevB.64.024512,doi:10.1143/JPSJ.69.2723,Mizusaki2002}.
The variance extrapolation works better for the RBM+PP than the P-RBM wave function because the variance is already small.
The accuracy of the extrapolated energy 
for  RBM+PP wave functions with spin and momentum quantum projections  $\mathcal{NL}^{K=0}  \mathcal{L}^{S=0} {\mathcal P}_{\rm G}^{\infty} | \phi_{\rm pair} \rangle $
(red line in Fig.~\ref{fig_Spin1}(b)) reaches 
an order of 
$10^{-5}$ (0.001 percent) in the relative error, which is comparable to the size of error bars of SSE-QMC calculations~\cite{PhysRevB.56.11678}.  
 
RBM+PP can also be used to accurately calculate other physical quantities, besides the energy. For instance,
 we also measure spin structure factor  $S({\bf q}) = \frac{1}{3N_{\rm site}} \sum_{i,j} \sum_{\alpha=x,y,z} \langle  S_i^{\alpha}  S_j^{\alpha} \rangle e^{i {\bf q} \cdot { ({\bf r}_i - {\bf r}_j })}$.
The result for $S({\bf q_{\rm peak}}) = S(\pi,\pi)$ is shown in Table \ref{Tab_spin}, indicating high accuracy of the correlation functions by the RBM+PP.


\subsection{Hubbard model}
\begin{figure}[tb]
\vspace{0.0cm}
\begin{center}
\includegraphics[width=0.49\textwidth]{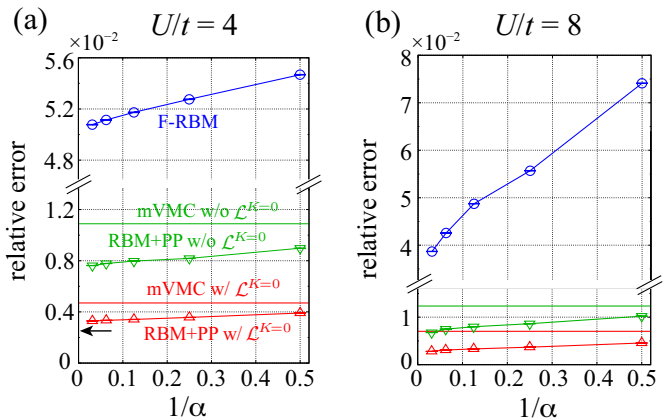}
\caption{
RBM+PP result  for energy of 2D Hubbard model at half filling with (a) $U/t=4$ and (b) $U/t=8$ on $8\times 8$ square lattice with 
P-AP 
boundary condition.  
Relative error of the RBM+PP energy to the AF-QMC energy ($E/t = -0.8642(2)$ and $-0.5259(3)$ for $U/t=4$ and $U/t=8$, respectively) \cite{PhysRevB.94.085103} as a function of $1/\alpha$ ($\alpha$: hidden variable density) is shown.
$\bigtriangleup$ ($\bigtriangledown$) symbol:
RBM+PP $| \Psi \rangle = \mathcal{NL}^{K=0}   | \phi_{\rm pair} \rangle $ ($| \Psi \rangle = \mathcal{N} | \phi_{\rm pair} \rangle$). 
$\bigcirc$ symbol: F-RBM $| \Psi \rangle = \mathcal{N} | \phi_{\rm Fermi\mathchar`-sea} \rangle$. 
Solid horizontal red (green) lines: results of mVMC functions $| \Psi \rangle = \mathcal{P}_{\rm G} \mathcal{P}_{\rm J} \mathcal{L}^{K=0}  | \phi_{\rm pair} \rangle $ ($| \Psi \rangle = \mathcal{P}_{\rm G} \mathcal{P}_{\rm J} | \phi_{\rm pair} \rangle$).
The arrow in (a) indicates the TNVMC result~\cite{TNVMC}.
Error bars show standard errors of Monte Carlo measurements of energy for the optimized variational wave function.  
}
\label{fig_Hubbbard}
\end{center}
\end{figure}

Figures.~\ref{fig_Hubbbard}(a) and  \ref{fig_Hubbbard}(b) show the RBM+PP result for the ground state energy of  $8\times8$ Hubbard model at half filling for $U/t=4$ and 8, respectively. 
The relative error to the auxiliary-field quantum Monte Carlo  (AF-QMC), which gives practically the exact results within the error bars, is plotted. 
For comparison, the mVMC results using Gutzwiller-Jastrow correlation factors and the F-RBM results are shown as well. 

In both cases ($U/t=4$ and $U/t=8$), the F-RBM has an error of several percent. 
The RBM+PP method, in which the variational parameters in both the $\mathcal N$ and $|\phi_{\rm pair}\rangle$ are optimized, achieves significantly higher accuracy. 
We see the same trend as the Heisenberg model; namely, the accuracy is improved by choosing a better reference wave function $| \phi_{\rm ref} \rangle $.

The RBM+PP wave function also surpasses the accuracy of the mVMC wave function, which indicates superiority of more unbiased neural-network factor ${\mathcal N}(x)$ to the empirical Gutzwiller-Jastrow factors ${\mathcal P}(x) =\langle x | \mathcal{P}_{\rm G} \mathcal{P}_{\rm J} | x \rangle $.
We expect that the advantage of a self-optimized neural network  will be more substantial for more complicated Hamiltonians than the single-band Hubbard model. In more complex Hamiltonians, more flexible forms for correlation factors are likely to be necessary.
Another advantage of the RBM+PP to the mVMC methods is that the accuracy improves as $U$ increases (see Appendix \ref{U-dep} and Fig.~\ref{fig_Udep}),  
while the mVMC results show the opposite trend~\cite{doi:10.1143/JPSJ.77.114701}. 

\begin{table} [tb]
\caption{ 
RBM+PP ($\mathcal{NL}^{K=0}   | \phi_{\rm pair} \rangle $) results for spin structure factor  $S(\pi,\pi)$ computed for 2D Hubbard model at half filling. 
For comparison, mVMC  ($\mathcal{P}_{\rm G} \mathcal{P}_{\rm J} \mathcal{L}^{K=0}  | \phi_{\rm pair} \rangle$) and AF-QMC~\cite{zhang_private} results are also listed.
}
\vspace{-0.5cm}
\begin{center}
\begin{tabular}{c  @{\ \  \  } c@{\ \  \ }  c   @{\ \   \ } c @{ \  \  } c   @{\   \ } c }
\hline 
\hline
\multirow{2}{*}{  } &    \multicolumn{5}{c}{   $S(\pi,\pi)/N_{\rm site} \times 10^2 $ }  \\
\cline{2-6}
   & $\alpha= 2$ & $\alpha= 8$ & $\alpha= 32$  & mVMC  & AF-QMC\\
\hline
$U/t \! =\! 4$  & \ 3.078(5)  & 3.057(5) & 3.021(5) & 3.107(4) & 2.92(2) \\
$U/t \! =\! 8$  & \ 5.233(9)  &5.206(9) & 5.20(1) & 5.30(1)    &  5.0(1) \\
\hline
\hline
\end{tabular}
\end{center}
\label{Tab_Hubbard}
\end{table}

At $U/t=4$ (Fig.~\ref{fig_Hubbbard}(a)), 
the best RBM+PP accuracy is comparable to the 
accuracy of TNVMC (tensor network combined with mVMC), where the relative error is $\sim 0.25$ percent as obtained at available maximum tensor bond dimension $D=16$, which is the computationally practical upper limit~\cite{TNVMC}.  
In the TNVMC method, the mVMC wave function is supplemented by the tensor network enabling one of the most accurate schemes among existing numerical methods~\cite{TNVMC}. 
Compared to the TNVMC method, the RBM+PP has an advantage, because it can be applied flexibly and easily to any kind of lattice and does not require an involved contraction procedure~\cite{ORUS2014117} in contrast to the TNVMC method. 

The spin structure factor $S(\pi,\pi)$ is listed in Table ~\ref{Tab_Hubbard}.
At both $U/t=4$ and $U/t=8$, with increasing $\alpha$, the value becomes closer to the exact AF-QMC value.


\section{Discussion}
\label{sec_discussion}
While the physical properties of the RBM have only recently started being discussed in condensed matter physics~\cite{PhysRevX.7.021021,Deng_arXiv,Levine_arXiv,Gao_Ncom,Chen_arXiv,Huang_arXiv,Gan_arXiv,Cai_arXiv,Glasser_arXiv}, 
more general discussion of representability can be traced back to 
earlier studies~\cite{HORNIK1991251,Cybenko1989,doi:10.1162/neco.2008.04-07-510}, which show that the RBM is able to describe any smooth function, 
if arbitrarily large $\alpha$ is allowed.
In the present 2D Heisenberg model, a gauge transformation can make
probability amplitude of the exact ground-state wave function $\ket{\Psi_{\rm GS}}$ positive ($\langle x | \Psi_{\rm GS} \rangle > 0$ for all $x$).
Therefore, the exact ground state can be represented by real-variable RBM with infinite $\alpha$. 
%
Accordingly, the relative error should go to zero as $1/\alpha \rightarrow 0$. 
Indeed, in Fig.~\ref{fig_Spin1}(a), it looks that the RBM+PP energy curves (red and green) start bending toward 0 as $1/\alpha$ decreases.  
It is likely that a better reference state makes $\langle x | \Psi_{\rm GS} \rangle / \langle x | \phi_{\rm ref}\rangle$ (to be represented by $\mathcal{N}(x)$) smoother and helps to reach
 faster convergence at small $\alpha$. 

For fermionic problems, nodal structure of wave functions is crucial~\cite{PhysRevLett.45.566}, which is beyond the representability of the real-variable RBM giving positive $\mathcal{N}(x)$~\cite{Cai_arXiv}.
In the RBM+PP, $|\phi_{\rm pair}\rangle$ is expected to accurately reproduce the nodal structure.
Then, $\langle x | \Psi_{\rm GS} \rangle / \langle x | \phi_{\rm pair}\rangle$ may become smooth enough so that with moderate $\alpha$, $\mathcal{N}(x)$ can represent a quick convergence to the exact value at $\alpha \rightarrow \infty$.
However, rigorously speaking, the nodal structure of  $|\phi_{\rm pair}\rangle$ is likely to be different from the exact one even when the variational parameters contained in $|\phi_{\rm pair}\rangle$ are ideally optimized.  
Therefore, introduction of complex variational parameters in the RBM part may be useful to adjust the nodes to the exact positions
beyond the framework of the Pfaffian wave function.
Although it is an interesting open question whether the energy curve as a function of $1/\alpha$ in the Hubbard model (Fig.~\ref{fig_Hubbbard}) goes to 0 as $1/\alpha \rightarrow 0$, 
in practical computations, the optimization at larger $\alpha$ becomes more and more difficult, which might hamper the expected convergence. 

\section{Summary and Perspectives}
\label{sec_summary}
In this work, we have proposed a new variational ansatz for studying the ground states of many-body interacting quantum systems. 
Our variational wave function, which we call RBM+PP, combines the RBM based on the ANN and the mVMC methods. 
We have shown that, in both 2D Heisenberg and Hubbard models, the RBM+PP results show a dramatic improvement of accuracy over simple neural network wave functions (the P-RBM and F-RBM wave functions). We also see the superiority of the RBM+PP to the mVMC method. 
Since the RBM+PP method can be flexibly applied not only to bosonic (or spin) systems but also to fermionic problems, the RBM+PP method offers a wide range of applications with high accuracy and a reasonable computational cost. 

As a future perspective, it would be interesting to go beyond the RBM structure  and introduce second hidden layer (deep Boltzmann machine (DBM)).
DBM is argued to have more efficient representation of certain many-body wave-functions than RBM~\cite{Gao_Ncom}.
In DBM, the spin variables of the neurons distributed in more than one hidden layer cannot be traced out analytically, thus we need to introduce additional Monte Carlo samplings for hidden spins.

\acknowledgements
We acknowledge useful discussions with Giuseppe Carleo, Takahiro Ohgoe, Hui-Hai Zhao, and Kota Ido. 
We also thank Shiwei Zhang for providing us with the auxiliary-field quantum Monte Carlo result of energy and spin structure factor in the 2D Hubbard model. 
The implementation of the RBM+PP scheme is done based on the mVMC package~\cite{mVMC_package}.
The computation was mainly done at Supercomputer Center, Institute for Solid State Physics, University of Tokyo.
This work was financially supported by Grant-in-Aids
for Scientific Research (JSPS KAKENHI) (No. 16H06345
and No. 17K14336) from Ministry of Education, Culture,
Sports, Science and Technology (MEXT), Japan. This work was
also supported in part by MEXT as a social and scientific
priority issue (Creation of new functional devices and
high-performance materials to support next-generation
industries (CDMSI)) to be tackled by using post-K computer. We also thank the support by the RIKEN Advanced Institute for Computational Science through
the HPCI System Research project (hp160201,hp170263) supported by MEXT.
Y. Y. was also supported by PRESTO, JST (JPMJPR15NF). 

\appendix
\setcounter{secnumdepth}{1}

\section{Ability of RBM to represent Gutzwiller and Jastrow factors}  
\label{Sec.RBM_Gutzwiller}
The Gutzwiller factor 
$\mathcal{P}_{\rm G} = \exp \large( - g  n_{i \uparrow} n_{i \downarrow} \large )$ 
at site $i$ can be recast as (except for trivial constant factor and one-body potential)
\begin{eqnarray}
\mathcal{P}_{\rm G} = \exp \left( - \frac{g}{4}  \sigma_{2i} \sigma_{2i-1}  \right )   
\end{eqnarray}
where $(\sigma_{2i},\sigma_{2i-1}) = (2n_{i\uparrow} \!-\!1,2n_{i\downarrow} \!-\!1)$ are physical variables in the visible layer defined in Sec.~\ref{sec_RBM_wf}.
This interaction between physical variables can be mediated by adding one hidden neuron $h$ as 
(except for trivial constant factor)
\begin{eqnarray}
\mathcal{P}_{\rm G} &=&  \sum_{h = \pm1} \exp \large ( W_1  \sigma_{2i}  h + W_2 \sigma_{2i-1}  h \large )    \nonumber \\
  &=& 2 \cosh \large( W_1  \sigma_{2i} + W_2  \sigma_{2i-1} \large)
\end{eqnarray}
with $W_1 \!  =  \!  \frac{1}{2} {\rm arcosh} \left( \exp \left( | g  |/2 \right ) \right)$ and 
$W_2 \!  = \! - {\rm sgn} \ g   \times   W_1$. 
This form is consistent with the neural-network factor ${\mathcal N}$ defined in Eq. (\ref{eq:Psi(x)}).
In the very same way, we can show that the neural-network correlation factor ${\mathcal N}$ can represent the Jastrow factor  
$ \mathcal{P}_{\rm J} = \exp \large(  -\frac{1}{2}\sum_{i,j  (i\neq j)} v_{ij} n_{i} n_{j} \large)  $.

\section{List of variational wave functions employed in simulations}  
\label{ListVF}
Tables~\ref{Table_Heisenberg} and \ref{Table_Hubbard} summarize the forms of the wave functions employed in the present study for solving the Heisenberg and Hubbard models, respectively.

\begin{table*} [tb]
\caption{ 
List of wave functions used in the analysis of the $8\times 8$  Heisenberg model ($N_{\rm site} = 64$). 
To describe the singlet state, the $\sigma$ and $\sigma'$ spins in the $f_{ij}^{\sigma \sigma'}$ parameters in the PP wave function 
are always set to be a pair of $\uparrow$ and $\downarrow$ spins.
We impose $2 \times 2$ sublattice structure in $f_{ij}^{\uparrow \downarrow}$ parameters in the PP (geminal) wave function. 
In this case, the number of independent $f_{ij}^{\uparrow \downarrow} $ parameters becomes $2 \times 2 \times N_{\rm site} = 4N_{\rm site}$, and the other 
 $f_{ij}^{\uparrow \downarrow} $ parameters are determined by using spatial translation operations.  
 The RBM part  ($\{  b_k, \ W_{ik} \}$) is taken to be fully translationally invariant ($1 \times 1$ sublattice structure)~\cite{Carleo602}. 
 When the double occupancy is completely prohibited by the Gutzwiller factor ${\mathcal P}_{\rm G}^{\infty} = \prod_i (1- n_{i \uparrow} n_{i \downarrow})$, 
 the onsite $f_{ii}^{\uparrow \downarrow}$ parameters become completely irrelevant, i.e., the wave function and the energy do not depend on  $f_{ii}^{\uparrow \downarrow}$  at all. 
 Thus the number of $f_{ij}^{\uparrow \downarrow}$ parameters in the table are reduced from $4N_{\rm site}$ to $4 (N_{\rm site} \! - \! 1)$.
 }
\vspace{0.0cm}
\begin{center}
\begin{tabular}{c@{  } c@{   } c@{   } c@{\   }  c    }
\hline 
\hline
          method \vspace{0.04cm}
      &    wave function
      &    symbol  in Fig. 2  
      &   variational parameters 
      & \# of variational parameters   \\
 \hline 
 P-RBM    \vspace{0.04cm}
&$ \mathcal{N}  | \phi_{\rm product} \rangle $   
& open circle (blue)
& $\{  b_k, \ W_{ik} \}$
&   $\alpha (N_{\rm site} \!+\!1)  = 65 \alpha$
 \\
RBM+PP    \vspace{0.04cm}
& $ \mathcal{N} \mathcal{L}^{S=0}  {\mathcal P}_{\rm G}^{\infty} | \phi_{\rm pair} \rangle $ 
 & down-pointing triangle (green)
 & $\{  b_k, \ W_{ik}, \   f_{ij}^{\uparrow \downarrow}\}$  
 & $\alpha (N_{\rm site}\!+\!1) \!+ \!4 (N_{\rm site}   \! - \!1 )= 65 \alpha \!+ \! 252$
 \\
 RBM+PP  \vspace{0.04cm}
& $ \mathcal{NL}^{K=0}  \mathcal{L}^{S=0}  {\mathcal P}_{\rm G}^{\infty} | \phi_{\rm pair} \rangle $ 
 &  up-pointing triangle (red)
 & $\{  b_k, \ W_{ik},   \ f_{ij}^{\uparrow \downarrow}\}$ 
 & $\alpha (N_{\rm site}\!+\!1) \!+\! 4 (N_{\rm site} \! - \! 1) = 65 \alpha \! + \!  252$
 \\ 
  mVMC \vspace{0.04cm}
& $  \mathcal{L}^{S=0}  {\mathcal P}_{\rm G}^{\infty} | \phi_{\rm pair} \rangle  $ 
 & green solid horizontal line
 & $\{  f_{ij}^{\uparrow \downarrow}\}$  
 & $ 4 (N_{\rm site} \! - \! 1) =  252$
 \\
 mVMC \vspace{0.04cm}
& $  \mathcal{L}^{K=0}  \mathcal{L}^{S=0}  {\mathcal P}_{\rm G}^{\infty} | \phi_{\rm pair} \rangle  $ 
 & red solid horizontal line
 & $\{    f_{ij}^{\uparrow \downarrow}\}$ 
 & $4 (N_{\rm site} \! - \! 1) =  252$
 \\ 
\hline
\hline
\end{tabular}
\end{center}
\label{Table_Heisenberg}
\end{table*}

\begin{table*} [tb]
\caption{ 
List of wave functions used in the analysis of the $8\times 8$ Hubbard model ($N_{\rm site} = 64$). 
As in the case of Heisenberg model, the $\sigma$ and $\sigma'$ spins in the $f_{ij}^{\sigma \sigma'}$ parameters are set to be always anti-parallel. 
In the mVMC method, we use Gutzwiller $\mathcal{P}_{\rm G}$ and Jastrow $\mathcal{P}_{\rm J}$ factors, whose forms are 
 $ \displaystyle \mathcal{P}_{\rm G} = \exp \left( -\sum_{i} g_i n_{i \uparrow} n_{i \downarrow} \right) $ and
$ \displaystyle \mathcal{P}_{\rm J} = \exp \left( -\frac{1}{2}\sum_{i,j  (i\neq j)} v_{ij} n_{i} n_{j} \right) $, respectively. 
All the $g_i$, $v_{ij}$, and $f_{ij}^{\uparrow\downarrow}$ parameters are taken to be independent ($8 \times 8$ or full sublattice structure). 
On the other hand, the RBM part is taken to be translationally invariant ($1 \times 1$ sublattice structure)~\cite{Carleo602} to save computational cost. 
We have confirmed that, in the $4\times4$ Hubbard model,  taking full sublattice structure in the RBM part does not help much to lower the energy 
compared to $1\times 1$ sublattice case, although it drastically increases the number of variational parameters and hence increases the computational cost. 
 }
\vspace{0.0cm}
\begin{center}
\begin{tabular}{c@{  \ } c@{ \    } c@{\ \   \   } c@{\ \ \   }  c }
\hline 
\hline
       method \vspace{0.04cm}
      &    wave function
      &    symbol  in Fig. 3  
      &   variational parameters 
      & \# of variational parameters   \\
\hline 
F-RBM \vspace{0.04cm}
&$ \mathcal{N}  | \phi_{\rm Fermi\mathchar`-sea} \rangle $   
& open circle (blue)
& $\{  b_k, \ W_{ik} \}$
&   $\alpha (2N_{\rm site} \!+\!1)  = 129 \alpha$
 \\
RBM+PP   \vspace{0.04cm}
& $ \mathcal{N}  | \phi_{\rm pair} \rangle $ 
 & down-pointing triangle (green)
 & $\{  b_k, \ W_{ik}, \   f_{ij}^{\uparrow \downarrow}\}$  
 & $\alpha (2N_{\rm site}\!+\!1) \!+\! 64 N_{\rm site}   = 129 \alpha + 4096$
 \\
 RBM+PP \vspace{0.04cm}
& $ \mathcal{NL}^{K=0}  | \phi_{\rm pair} \rangle $ 
 & up-pointing triangle (red)
 & $\{  b_k, \ W_{ik},   \ f_{ij}^{\uparrow \downarrow}\}$ 
 & $\alpha (2N_{\rm site}\!+\!1) \!+\! 64 N_{\rm site} = 129 \alpha + 4096$
 \\
 mVMC \vspace{0.04cm}
& $ \mathcal{P}_{\rm G} \mathcal{P}_{\rm J}  | \phi_{\rm pair} \rangle $ 
 & green solid horizontal line
 & $\{  g_i, \  v_{ij}, \   f_{ij}^{\uparrow \downarrow}\}$  
 & $ N_{\rm site} \!+ \! 32 (N_{\rm site} \!- \! 1) \!+\! 64 N_{\rm site}   = 6176$
 \\
 mVMC \vspace{0.04cm}
& $ \mathcal{P}_{\rm G} \mathcal{P}_{\rm J} \mathcal{L}^{K=0}  | \phi_{\rm pair} \rangle $ 
 & red solid horizontal line
 & $\{  g_i, \  v_{ij},   \ f_{ij}^{\uparrow \downarrow}\}$ 
 & $ N_{\rm site} \!+\! 32 (N_{\rm site} \!- \! 1) \!+\! 64 N_{\rm site} = 6176$
 \\ 
\hline
\hline
\end{tabular}
\end{center}
\label{Table_Hubbard}
\end{table*}

\section{Details of calculation conditions}
\label{Details}
\setcounter{secnumdepth}{2}



\subsection{Stabilization factor}
In the present study, the parameters are optimized by the stochastic reconfiguration (SR) method~\cite{PhysRevB.64.024512}. 
The same optimization scheme is called natural gradient in Refs.~\cite{IEEE_Amari,neural_Amari}.
This optimization is equivalent to the imaginary-time evolution $e^{-\tau \mathcal{H} }|\Psi\rangle$ of the wave function $|\Psi\rangle$ to reach the ground state for sufficiently large imaginary time $\tau$ in the truncated Hilbert space which is spanned by the variational wave function. 
In the SR optimization, the variational parameters $\gamma_{m}$ $(m=1,2,\ldots, N_{\rm v})$ at the $p$-th iteration are updated as 
\begin{eqnarray}
 \gamma_{m}^{(p+1)} = \gamma_{m}^{(p)}   + \Delta \gamma_{m}^{(p)}, 
\end{eqnarray}
where the difference in the update $\Delta \gamma_{m}^{(p)}$  is given by 
\begin{eqnarray}
   \sum_{n=1} ^{N_{\rm v}}  S_{mn}^{(p)}    \Delta  \gamma_n^{(p)} =  - \Delta \tau   \ \! g_{m} ^{(p)}.
\end{eqnarray}
with a small imaginary time step $\Delta \tau$. 
Here, $S$ is a positive definite matrix given by 
\begin{eqnarray}
S_{mn} =  \langle  \partial_{\gamma_m}  \bar{\Psi}  |  \partial_{\gamma_n} \bar{\Psi} \rangle 
\end{eqnarray}
with a normalized variational wave function $| \bar{\Psi} \rangle =  | \Psi \rangle  / \sqrt{ \langle \Psi | \Psi \rangle }$ 
and $ |  \partial_{\gamma_m} \bar{\Psi} \rangle  = \frac{\partial}{\partial \gamma_m} | \bar{\Psi} \rangle$.
The $g$ vector is
 the gradient of energy with respect to the $\gamma$ parameters: 
\begin{eqnarray}
 g_{m} = \frac{\partial E}{ \partial \gamma_m}  =   \frac{\partial }{ \partial \gamma_m}   \langle   \bar{\Psi}  |  {\mathcal H}|   \bar{\Psi} \rangle 
\end{eqnarray}

To stabilize the optimization, we introduce the stabilization factor to the diagonal elements of the $S$ matrix as 
\begin{eqnarray}
  S_{mm}^{(p)}  \rightarrow  S_{mm}^{(p)} ( 1 +  \epsilon_1^{(p)} )  + \epsilon_2^{(p)}
\max_m \left\{ S_{mm}^{(p)}  \right \}.
\end{eqnarray}
Here, $ \epsilon_1$ scales the diagonal elements and $\epsilon_2$ gives a constant shift to the $S$ matrix. 
Although $\epsilon_2$ strongly stabilize the optimization, it sometimes makes convergence slower, so we typically put a small number to $\epsilon_2$: 
$\epsilon_2^{(p=0)} \sim 10^{-3}$ and  we gradually decrease to $\epsilon_2^{\infty} \sim 10 ^{-7}$-$10^{-6}$ through the first several hundred iterations of optimization.  
As for $\epsilon_1$, we found that smaller $\epsilon_1 \ (\sim 10^{-5} \mathchar`- 10^{-4}) $ factor for neural-network-related variational parameters $\{ b_k, \ W_{ik} \}$  sometimes helps to lower the energy. 
Meanwhile, we also found that a small $\epsilon_1 \ (\lesssim 10^{-3})$ makes the optimization of $f_{ij}^{\uparrow\downarrow}$ parameters in the pair-product (geminal) wave function unstable.
To overcome this problem, we use parameter dependent $\epsilon_1 $ factor, namely, 
$\epsilon_1^{(p)} (\{ f_{ij}^{\uparrow\downarrow} \} ) =  \epsilon_1^{(p)} (\{  b_k, \ W_{ik} \} ) + \Delta \epsilon_1$ with $\Delta \epsilon_1 \sim 10^{-2}$.   
With this condition, we typically use $\epsilon_1^{(p=0)} (\{  b_k, \ W_{ik} \} )  \sim 10^{-2} \mathchar `-  10^{2}$ 
and gradually decrease 
it to $\epsilon_1^{\infty} (\{  b_k, \ W_{ik} \} ) \sim 10^{-5} \mathchar`- 10^{-4}$ in the first several hundred iterations of the optimization. 

As for the initial  $\{  b_k, \ W_{ik} \}$ parameters, we use small random numbers. 
We run several calculations with different seeds for generating random numbers and 
adopt the wave function that has the lowest energy. 

\subsection{Particle-hole transformation of Hubbard model}
When we analyze many-body Hamiltonians by theoretical or numerical solvers,
we can utilize transformations of the Hamiltonians
to find a representation suitable for the solver in hand~\cite{doi:10.1143/JPSJ.57.2482}.
In the conventional variational Monte Carlo method, many-body interacting models defined in the real-space basis such as the Hubbard model is analyzed as they are, 
because the empirical form of the correlation factors is also defined in the real-space basis. 
On the other hand, in cases where we employ machine learning technique,  because the neural network will find a way to lower the energy even when the form of the Hamiltonian is complicated, we could think of the ``best" transformation of the Hamiltonian such that the neural network can lower the energy with small number of parameters. 

Though it would not be the best of the best, we find that performing a staggered particle-hole transformation ($c_{i \downarrow} \rightarrow  (-1)^{i} c^{\dagger}_{i \downarrow}$  and $c^{\dagger}_{i \downarrow} \rightarrow  (-1)^{i} c_{i \downarrow}$)
and mapping onto the attractive Hubbard model help the RBM+PP wave function to lower the energy 
of the Hubbard model. 
This transformation is also helpful for mVMC calculations to get a better energy~\cite{ohgoe_private}.
Thus, we have solved the Hubbard model with this transformation. 
 Finding a better transformation (by again employing machine learning technique) is an interesting future problem.

\section{$U$ dependence of energy in RBM+PP }
\label{U-dep}

\begin{figure}[tb]
\vspace{0cm}
\begin{center}
\includegraphics[width=0.38\textwidth]{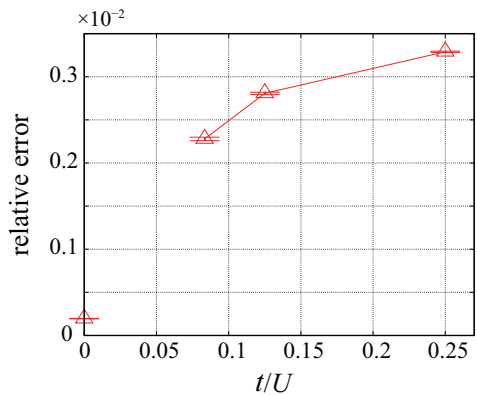}
\caption{
$U$ dependence of RBM+PP ($\mathcal{NL}^{K=0}   | \phi_{\rm pair} \rangle $ with $\alpha = 32$) 
energy of 2D Hubbard model defined on $8\times 8$ square lattice with 
P-AP boundary condition.  
Relative error of the RBM+PP energy to that obtained by the AF-QMC \cite{PhysRevB.94.085103,zhang_private} 
is shown. For comparison, at $t/U=0$, we show the RBM+PP ($ \mathcal{NL}^{K=0}  \mathcal{L}^{S=0}  {\mathcal P}_{\rm G}^{\infty} | \phi_{\rm pair} \rangle $ with $\alpha=32$) result for the $8\times 8$ Heisenberg model. 
Error bars show standard errors of Monte Carlo measurements of energy for the optimized variational wave function. 
}
\label{fig_Udep}
\end{center}
\end{figure}

Figure~\ref{fig_Udep} shows the 
$U$ dependence of RBM+PP ($ \mathcal{NL}^{K=0}   | \phi_{\rm pair} \rangle $ with $\alpha = 32$) 
energy of the 2D $8\times 8$ Hubbard model.
The result shows that the error decreases with increasing $U/t$.

\bibliographystyle{apsrev}
\bibliography{main}


\end{document}